\documentclass[12pt]{article}
\usepackage{amssymb}
\usepackage{amsmath}
\usepackage{graphicx}
\usepackage{nicefrac}
\title{A scaling invariance of the perturbations in $k$-inflation models}
\author{
	Neven Bili\'c\thanks{bilic@irb.hr}
	\\
	{\it Division of Theoretical Physics, Rudjer Bo\v{s}kovi\'{c} Institute,}
	\\ {\it Zagreb 10000, Croatia}
	\vspace{0.1in}
	\and
	Dragoljub D.\ Dimitrijevi\'{c}\thanks{ddrag@pmf.ni.ac.rs}, 
	Goran S.\ Djordjevic\thanks{gorandj@junis.ni.ac.rs}, Milan Milo\v{s}evi\'{c}\thanks{milan.milosevic@pmf.edu.rs}
	\\
	{\it Department of Physics,
		University of Ni\v{s}, Ni\v{s} 18000,  Serbia}
	\vspace{0.1in}
	\and
	Marko Stojanovi\'{c}\thanks{marko.stojanovic@pmf.edu.rs} 
	\\
	{\it Faculty of Medicine, University of Ni\v{s}, Ni\v{s} 18000, Serbia}
}

\begin{document}
\maketitle
\flushbottom

	\thispagestyle{empty}
	
	\vspace{0.3cm}

\abstract{
We study the background and perturbations in $k$-essence inflation models and show that a general $k$-essence exhibits a simple scaling property.
In particular, we study two classes of $k$-inflation models
with the potential characterized by an inflection point.
We demonstrate that these models enjoy scaling properties that could be used to redefine input parameters so that the perturbation spectra
satisfy correct normalization at the CMB pivot scale. The background and perturbation equations are integrated numerically for two specific models.
}

\section{Introduction}

The origin of the field driving inflation is still unknown and subject to speculation. The models in which the kinetic term has a noncanonical form are usually
referred to as $k$-essence \cite{armendariz2} or, in the context of inflation, $k$-inflation models \cite{armendariz1}.  
As argued by  Armend\'ariz-Pic\'on,  Damour,  and Mukhanov
\cite{armendariz1}, a field theory with non-canonical kinetic terms can be 
motivated by a low-energy realization of string theory.
Suppose we have only gravity and
some moduli field $\varphi$ (e.g., the dilaton) in
string theory. Then, $\alpha'$ corrections (due to the massive modes
of the string) generate a series of higher-derivative terms in the low-energy
effective action, while string-loop corrections generate a non-trivial 
moduli dependence of the coefficients of the various kinetic terms.
Passing from the string to the Einstein frame, one obtains the effective Lagrangian of the form
$\mathcal{L}=K(\varphi) X+L(\varphi) X^2 +\cdots$, where $X=g^{\mu\nu}\varphi_{,\mu}\varphi_{,\nu}$.
More details can be found in Ref. \cite{armendariz1}.

A distinguished feature of $k$-essence models is that even without any potential
energy term, a general class of non-standard kinetic-energy terms can drive an inflationary evolution
of the same type as the usually considered potential-driven inflation.
For example, a simple $k$-essence model with Lagrangian $\mathcal{L}= -A\sqrt{1-X}$ in the slow-roll regime produces
an accelerated expansion, i.e., 
a quasi-de Sitter evolution. In the cosmological context, this model is equivalent to the so-called Chaplygin gas 
\cite{kamenshchik} and serves as
a prototype model for the dark matter/dark energy unification \cite{bilic}.
Promoting the constant $A$ to a field-dependent potential, this model becomes
the Tachyon model -- a particularly important $k$-essence model extensively studied in the literature 
\cite{fairbairn,frolov,shiu1,sami,shiu2,kofman,cline,salamate2018,barbosa2018,dantas2018,steer,bilicJCAP}.

In this work, we study the $k$-essence using the Hamiltonian formalism for the background field equations. 
We show that a general $k$-essence exhibits a simple  scaling property
under the condition that the Lagrangian admits a multiplication by a constant.
We then specify our studies to two broad classes of $k$-essence.

In class  A,
the Lagrangian is of the form $(X/2)^\alpha -U(\varphi)$,
where $\alpha\geq 1$ is a constant, $X=g^{\mu\nu}\varphi_{,\mu}\varphi_{,\nu}$, and  $U$ is a smooth function of the field $\varphi$. A typical representative of this class is the model proposed by Papanikolaou, Lymperis, Lola, and Saridakis \cite{papanikolaou}, which
we will refer to as the PLLS model.   
Their potential is a function of the field with an inflection point.
The potentials with inflection or near inflection point always lead to a peak in the power spectrum of curvature perturbations on scales below the CMB pivot scale. The ehanced cosmological perturbations could collapse into primordial black holes (BHs) \cite{garcia-bellido,kannike,germani}.

Class B is represented by  
the Lagrangian of the form  $-U(\varphi)F(X)$, where $F$ is a smooth function of $X$.
A typical representative of this class is the Tachyon model with a Lagrangian of the Dirac-Born-Infeld (DBI) form 
\begin{equation}
\mathcal{L}_{\rm DBI}=-U(\varphi)\sqrt{1-X} .
\label{eq8001}
\end{equation}
Our interest in the Tachyon model is motivated mostly by low-energy string theory.
The existence of tachyons in the perturbative spectrum of string theory shows that the perturbative vacuum is unstable. This instability implies that there is a real vacuum to which a tachyon field $\varphi$ tends \cite{gibbons}.
The foundations of this process are given by a model of effective field theory \cite{sen,Sen:2002in}
with the Lagrangian of the DBI form.
A similar Lagrangian appears in the so-called brane inflation models \cite{shandera}. In these models, inflation is driven by the motion of a D3-brane in a warped throat region of a compact space, and the DBI field corresponds to the position of the D3-brane.
To see how the tachyon Lagrangian appears through the dynamics of a 3-brane 
(see Ref.\ \cite{bilic2} for details)
consider
a 3-brane 
moving in the 4+1-dimensional bulk spacetime 
with coordinates $X^{a}$,
$a=0,1,2,3,4$ and line element
 \begin{equation}
 	ds^2_{(5)}=G_{ab} dX^a dX^b=\chi(z)^2( g_{\mu\nu}dx^\mu  dx^\nu 
 -dz^2).
 \label{eq0002}
\end{equation}
The points on the brane are parameterized by
$X^a (x^{\mu})$, $\mu=0,1,2,3$, where $x^{\mu}$
are the coordinates on the brane.
The brane action is given by  
\begin{equation}
	S_{\rm br}= - \sigma
	\int d^4x\, \sqrt{-\det (G_{ab} X^a_{,\mu} X^b_{,\nu})}  \, , 
	\label{eq0001}
\end{equation}
where $\sigma$ is the brane tension.
Taking the Gaussian normal parameterization 
$X^a(x^\mu)=\left(x^\mu, z(x^\mu)\right)$,
a straightforward calculation of the determinant yields  
the brane  action Lagrangian in the form (\ref{eq8001})
in which the role of the field is played by the extra coordinate $z$ with $X=g^{\mu\nu} z_{,\mu} z_{,\nu}$ and $U(z)=\sigma \chi(z)^4$.

We will demonstrate that both A and B  models enjoy scaling properties which one can use to tune the model parameters and initial values 
without affecting the shape of the curvature perturbation spectrum.
The tuning of the input parameters is often necessary to achieve the desired properties of the perturbation spectrum but inevitably changes 
the normalization of the spectrum if one adheres to the Bunch-Davies asymptotic condition. However, owing to the scaling properties one can redefine the input parameters so that the spectrum keeps its shape and satisfies the correct normalization at the CMB pivot scale. 

The remainder of the paper is organized as follows. In Sec.\ \ref{hamilton}, we give an overview of the covariant Hamilton formalism. In Sec.\ \ref{k-essence}, we consider two classes of $k$-essence and present the formalism that describes the background evolution. 
In Section \ref{perturbations}, we investigate
the spectra of curvature perturbations for these two classes of models.
In the last section, Sec.\ \ref{conclude}, we summarize our main results and lay out conclusions. 

{\bf Notation}
\\
We use the metric signature $(+,-,-,-)$.
For convenience, we introduce a length scale $\ell$ so that $\ell\gg 1/M_{\rm Pl}$, where
$M_{\rm Pl}=\sqrt{1/(8\pi G)}$ is the reduced Planck mass.
By $\mathcal{L}$ and $\mathcal{H}$, we
denote the Lagrangian and Hamiltonian multiplied by $\ell^4$ so that our Lagrangian and Hamiltonian are dimensionless.
By $\varphi$ and $\phi$, we denote the scalar fields connected by $\varphi=\ell^2\phi$ and have the dimension of length and mass, respectively.


\section{Hamilton formalism}
\label{hamilton}
For a general (dimensionless) scalar field Lagrangian 
$\mathcal{L}(\varphi_{,\mu},\varphi)$,
the (dimensionless) covariant Hamiltonian ${\cal H}(\eta^\mu, \varphi)$
is related to
${\cal L}$ 
through the Legendre transformation
\begin{equation}
	{\cal H} (\eta^\mu, \varphi)= \eta^\mu\varphi_{,\mu} -{\cal L} (\varphi_{,\mu}, \varphi) ,
	\label{eq0024}
\end{equation}
with conjugate variables satisfying the conditions
\begin{equation}
	\varphi_{,\mu} =  \frac{\partial{\cal H}}{\partial\eta^{\mu}},
	\label{eq0025}
\end{equation}
\begin{equation}
	\eta^\mu =  \frac{\partial{\cal L}}{\partial\varphi_{,\mu}}.
	\label{eq0026}
\end{equation}
From (\ref{eq0024}) we have
\begin{equation}
	\frac{\partial{\cal H}}{\partial\varphi}=
	-\frac{\partial{\cal L}}{\partial\varphi}.
	\label{eq1026}
\end{equation}
By making use of Eqs. (\ref{eq0026}), (\ref{eq1026}),  and the Euler-Lagrange equation 
\begin{equation}
	\left(\frac{\partial{\cal{L}}}{\partial\varphi_{,\mu}}\right)_{;\mu}
	=\frac{\partial{\cal{L}}}{\partial\varphi} ,
	\label{eq0029}
\end{equation}
we find 
\begin{equation}
	{\eta^\mu}_{;\mu}=-\frac{\partial{\cal H}}{\partial\varphi} .
	\label{eq2026}
\end{equation}
Equations (\ref{eq0025}) and (\ref{eq2026}) are the covariant Hamilton equations. 

The most general scalar field Lagrangian is a function of the
form $\mathcal{L}= \mathcal{L}(X,\varphi)$,
where
\begin{equation}
	X=g^{\mu\nu}\varphi_{,\mu}\varphi_{,\nu}.
	\label{eq6001}
\end{equation}
This type of Lagrangian represents
the so-called $k$-essence theory.
In cosmological applications
the kinetic term $X$
is assumed positive.  
From (\ref{eq0026}) it follows that the quantity
\begin{equation}
	\eta^2=g_{\mu\nu}\eta^\mu\eta^\nu=4X(\mathcal{L}_X)^2 
	\label{eq6012}
\end{equation}
is also  positive.
Here, the subscript $_X$ denotes a derivative with respect to $X$.
Then, one can introduce the four-velocity
\begin{equation}
	u_\mu=\frac{g_{\mu\nu}\eta^\nu}{\eta}= \epsilon\frac{\varphi_{,\mu}}{\sqrt{X}},
	\label{eq0032}
\end{equation}
where, $\epsilon$ is $+1$  or $-1$ according to whether $\varphi_{,0}$ is respectively positive
or negative. This choice of $\epsilon$ guaranties that $u_0$ is always  positive.
The requirement of positivity of $u_0$ is natural in the context of FRW cosmology, where we require that 
the expansion rate 
\begin{equation}
H\equiv\frac{1}{3}{u^\mu}_{;\mu}
\end{equation} 
equals $\dot{a}/a$ in comoving frame.
The quantity $\eta$ in (\ref{eq0032}) is a square root of (\ref{eq6012}). The sign of the root is fixed by 
\begin{equation}
	\eta \equiv 2\epsilon\sqrt{X}{\cal L}_X= \epsilon \frac{\partial{\cal L}}{\partial\sqrt{X}},
	\label{eq6002}
\end{equation}
which follows from (\ref{eq6012}) and (\ref{eq0032}).
Here and throughout the paper, we assume  $\sqrt{X}\geq 0$.
Hence, $\eta$
can be positive or negative depending on the signs of $\epsilon$ and $\mathcal{L}_X$.

In a $k$-essence type of theory, 
the Hamiltonian is a function of 
$\eta^2=g_{\mu\nu}\eta^\mu\eta^\nu$ and $\varphi$. One can see this as follows.
First, using (\ref{eq0024}) and (\ref{eq0026}) one can express $\mathcal{H}$ as a function of $X$ and $\varphi$,  
\begin{equation}
	{\cal H} = 2X {\cal L}_X -{\cal L} .
	\label{eq6004}
\end{equation}
Then, from (\ref{eq6002}) it follows that 
$X$ is an implicit function of $\eta^2$ and $\varphi$ which we can 
(in principle) solve for $X$ to obtain
$X=X(\eta^2,\varphi)$. Then, we can plug in this solution
into the right-hand side of (\ref{eq6004}) to obtain  $\mathcal{H}= \mathcal{H}(\eta^2,\varphi)$.
Besides, one can show that the Lagrangian can be expressed as a function of
$\eta^2$ and $\varphi$ 
\begin{equation}
	\mathcal{L}=\eta\frac{\partial{\cal H}}{\partial\eta}-
	\mathcal{H},
	\label{eq0045}
\end{equation}
which follows from  (\ref{eq0024}), (\ref{eq0025}), and the definition
$\eta^2=g_{\mu\nu}\eta^\mu \eta^\nu$.

Using (\ref{eq0025})-(\ref{eq0029}) and  (\ref{eq6001})-(\ref{eq0032}) we can write the Hamilton equations in the standard form
\begin{equation}
	\dot{\varphi} =  \frac{\partial{\cal H}}{\partial\eta},
	\label{eq0033}
\end{equation}
\begin{equation}
	\dot{\eta} +3H \eta = - \frac{\partial{\cal H}}{\partial\varphi},
	\label{eq3033}
\end{equation}
where we have used the usual notation $\dot{f}\equiv u^\mu f_{,\mu} $
and the definition of the expansion rate.
As usual, we identify the Lagrangian and  Hamiltonian  with the pressure $p$ and energy density $\rho$,
respectively.
Then, in the cosmological context, 
the Hubble expansion rate and its time derivative can be related to the Hamiltonian and Lagrangian. Using the Friedmann equations, we find
\begin{equation}
	H^2=\frac{\mathcal{H}}{3\ell^4 M_{\rm Pl}^2} ,
	\label{eq3060}
\end{equation}
\begin{equation}
	\dot{H} = -\frac{\mathcal{L}+\mathcal{H}}{2\ell^4M_{\rm Pl}^2}
	=-\frac{\eta\mathcal{H}_{,\eta}}{2\ell^4M_{\rm Pl}^2}.
	\label{eq4020}
\end{equation}
It is sometimes convenient to express Eqs.\ (\ref{eq0033}) and (\ref{eq3033})
in terms of the Lagrangian. With the help of (\ref{eq1026}),(\ref{eq6012}), and (\ref{eq6004}) we find
\begin{equation}
	\dot{\varphi} =  \frac{\eta}{2\mathcal{L}_X},
	\label{eq4033}
\end{equation}
\begin{equation}
	\dot{\eta} +3H \eta =  \frac{\partial{\mathcal{L}}}{\partial\varphi}.
	\label{eq5033}
\end{equation}

It is worth noting that  
equation (\ref{eq0024}) can be written as a Legendre transformation 
\begin{equation}
	{\cal H} (\eta, \varphi)=  \eta \varphi_{,0}  -{\cal L} (\varphi_{,0} , \varphi) ,
	\label{eq0027}
\end{equation}
with two conjugate variables
$\varphi_{,0} \equiv\epsilon\sqrt{X}$ and $\eta$ satisfying 
\begin{equation}
	\varphi_{,0} 	 =  \frac{\partial{\cal H}}{\partial\eta},
	\label{eq0028}
\end{equation}
\begin{equation}
	\eta =  \frac{\partial{\cal L}}{\partial\varphi_{,0} }.
	\label{eq6030}
\end{equation}
Equation (\ref{eq0028}) is  the first Hamilton equation  equivalent to (\ref{eq0033}). The second 
Hamilton equation equivalent to (\ref{eq3033}) can be obtained from  the Euler-Lagrange equation
\begin{equation}
	\frac{1}{\sqrt{-g}}	\left(\sqrt{-g}\frac{\partial\mathcal{L}}{\partial\varphi_{,0} }\right)_{,0}=
	\frac{\partial\mathcal{L}}{\partial\varphi},
\end{equation} 
with (\ref{eq6030}) and (\ref{eq1026}).
Note also that Eq.\ (\ref{eq0045}) follows directly from (\ref{eq0027}) and (\ref{eq0028}).

\subsubsection*{A few useful relations} 

{\bf The speed of sound squared} can be expressed  either as a function 
of $X$ and $\varphi$,
\begin{equation}
	c_{\rm s}^2\equiv \left.\frac{\partial p}{\partial\rho}\right|_\varphi =\frac{\mathcal{L}_X}{\mathcal{H}_X}
	=\frac{\mathcal{L}_X}{\mathcal{L}_X+2X\mathcal{L}_{XX}},
	\label{eq3018}
\end{equation}
or as a function of $\eta$ and $\varphi$,
\begin{equation}
	c_{\rm s}^2 =\frac{\mathcal{L}_{,\eta}}{\mathcal{H}_{,\eta}}
	=\frac{\eta\mathcal{H}_{,\eta\eta}}{\mathcal{H}_{,\eta}} .
	\label{eq3019}
\end{equation}
Here, the subscripts ,$\eta$ and ,$\eta\eta$ denote respectively the first and second order partial derivatives with respect to $\eta$.
\\
\\
{\bf The first slow roll parameter} $\varepsilon_1$ can be expressed in terms of $\eta$ and $\varphi$ using the second Friedmann equation (\ref{eq4020}) yielding
\begin{equation}
	\varepsilon_1 \equiv -\frac{\dot{H}}{H^2}= \frac32
	\frac{\eta\mathcal{H}_{,\eta}}{\mathcal{H}}.
	\label{eq4021}
\end{equation}

\section{$k$-essence inflation models}
\label{k-essence}
Inflation models based on a scalar field theory with noncanonical 
kinetic terms are usually referred to as $k$-essence
\cite{armendariz2}.
For a general $k$-essence of the form $\mathcal{L}=\mathcal{L}(X,\lambda\varphi)$, where 
$\lambda$ is a parameter of dimension of mass,
one can demonstrate 
simple scaling properties
which can be useful for tuning the model parameters and initial values. 
First,  we introduce the e-fold number
\begin{equation}
	N=\int dtH,
	\label{eq3030}
\end{equation}
and rewrite the Hamilton equations (\ref{eq0033})  and (\ref{eq3033})
as differential equations with respect to $N$
\begin{equation}
	\frac{d\varphi}{dN}= \frac{1}{H} \frac{\partial{\cal H}}{\partial\eta},
	\label{eq6033}
\end{equation}
\begin{equation}
	\frac{d\eta}{dN}+3\eta=- \frac{1}{H}\frac{\partial{\cal H}}{\partial\varphi},
	\label{eq7033}
\end{equation} 
It may easily be shown that Eqs.\ (\ref{eq6033}) and (\ref{eq7033}) are invariant under the following simultaneous rescaling 
\begin{equation}
	\mathcal{L} \rightarrow c_0^{-1}\mathcal{L}, \quad
	\lambda \rightarrow c_0^{-1/2}\lambda, \nonumber
\end{equation}
\begin{equation}
	\varphi \rightarrow c_0^{1/2} \varphi , \quad
	\eta \rightarrow c_0^{-1} \eta ,	
	\label{eq90077}
\end{equation}
where $c_0$ is an arbitrary positive constant.

Next, we consider two popular classes of $k$-essence
described as follows.
\subsection{Model A}

This class of models is defined by the Lagrangian
\begin{equation}
	\mathcal{L}=
	\left(\frac{X}{2}\right)^\alpha-U(\lambda\varphi),
	\label{eq0035}
\end{equation}
where $\alpha >1$ is a dimensionless constant,
$\lambda>0$ is a constant of dimension of mass, and 
$\varphi$ is the field of dimension of length.
The physical field $\phi$,  related to $\varphi$ via
\begin{equation}
	\varphi=\ell^2\phi,
	\label{eq4036}
\end{equation}
has the usual dimension of mass. 
We assume that the kinetic term $X\equiv g^{\mu\nu}\varphi_{,\mu}\varphi_{,\nu}$ is positive and 
that
$U$ is a smooth dimensionless function of its argument. 
As described in Sec.\ \ref{hamilton}, the corresponding Hamiltonian is obtained by the Legendre transformation (\ref{eq6004}) 
yielding
\begin{equation}
	\mathcal{H}=	
	(2\alpha-1)\left(\frac{X}{2}\right)^{\alpha}+
	U(\lambda\varphi).
	\label{eq2035}
\end{equation}
Then, 
the Hubble expansion rate $H$ is given by (\ref{eq3060})
and the speed of sound by
\begin{equation}
	c_{\rm s}^2 \equiv 
	\frac{\mathcal{L}_X}{\mathcal{H}_X}
	=\frac{1}{2\alpha-1}.
\end{equation}

Now we introduce the conjugate momentum and its magnitude $\eta$ as described in 
Sec.\ \ref{hamilton}.
In this model, $X$ can be expressed as an explicit function of $\eta$.
By making use of (\ref{eq0035}) and (\ref{eq6002}) we find
\begin{equation}
	\frac{X}{2}=\left( \frac{\eta^2}{2\alpha^2}\right)^{1/(2\alpha-1)}.
	\label{eq0036}
\end{equation}
Then, we obtain the Hamiltonian expressed in terms of $\varphi$ and $\eta$ as
\begin{equation}
	\mathcal{H}=
	(2\alpha-1)\left( \frac{\eta^2}{2\alpha^2}\right)^{\alpha/(2\alpha-1)}+
	U(\lambda\varphi) .
	\label{eq0037}
\end{equation}
The first slow-roll parameter is  given by
\begin{equation}
	\varepsilon_1=
	\frac{\mathcal{L}+\mathcal{H}}{2\ell^4 M_{\rm Pl}^2H^2}=
	\frac{3\alpha\left[ \eta^2/(2\alpha^2)\right]^{\alpha/(2\alpha-1)}}{(2\alpha-1)\left[ \eta^2/(2\alpha^2)\right]^{\alpha/(2\alpha-1)}+
		U(\lambda\varphi)}.
\end{equation}
The end of inflation is determined by the
condition $1-\varepsilon_1\simeq 0$. 

The dynamics are governed by the Hamilton equations (\ref{eq6033})  and (\ref{eq7033}) defined 
for a general $k$-essence.
In this model, we find
\begin{equation}
	\frac{d\varphi}{dN} =\frac{\eta}{\alpha H}\left( \frac{\eta^2}{2\alpha^2}\right)^{(1-\alpha)/(2\alpha-1)} ,
	\label{eq7036}
\end{equation}
\begin{equation}
	\frac{d\eta}{dN}= -3\eta  - \frac{1}{H}\frac{\partial{U}}{\partial\varphi},
	\label{eq7034}
\end{equation}
where $ H$ is the Hubble rate 
\begin{equation}
	H=\frac{1}{\sqrt3\,\ell^2 M_{\rm Pl}}\left[	(2\alpha-1)\left(\frac{\eta^2}{2\alpha^2}\right)^{\alpha/(2\alpha-1)}
	+ U(\lambda\varphi)
	\right]^{1/2}.	 
	\label{eq7035}
\end{equation}

The scaling property of the general $k$-essence described by (\ref{eq90077}) does not apply 
to this model since we do not have the freedom to rescale the kinetic part.
Nevertheless, it may be shown that the above Hamilton equations  are invariant under the following simultaneous rescaling of the potential parameters and fields
\begin{equation}
	U \rightarrow c_0^{-1} 	U ,\quad
	\lambda \rightarrow c_0^{(1-\alpha)/(2\alpha)} \lambda, \nonumber 
\end{equation}
\begin{equation}
	\varphi \rightarrow c_0^{(\alpha-1)/(2\alpha) } \varphi , \quad
	\eta \rightarrow c_0^{(1-2\alpha)/(2\alpha)} \eta. 
	\label{eq8007}
\end{equation}
The Hubble rate scales according to
$H\rightarrow c_0^{-1/2} H$ 
and the rescaled Hamilton equations retain the form (\ref{eq7036})
and (\ref{eq7034}). 

Note that any canonical single-field inflation model enjoys the above scaling property.   On the other hand, in multifield inflation models, this type of scaling is not a general property. However, it is not excluded that some canonical multifield inflation models obey a similar scaling property, depending on the particulars of the potential.

\subsubsection*{The Klein-Gordon equation}
Instead of Hamilton's equations, one can use the second order Klein-Gordon (KG) equation, i.e., the field equation of motion. This equation can be obtained directly from the Lagrangian  or combining Eqs.\ (\ref{eq7036}) and (\ref{eq7034}). 
Either way one finds
\begin{equation}
	\varphi\,^{\prime\prime}+\left(\frac{3}{2\alpha-1}-
	\varepsilon_1\right) \varphi\,^{\prime}
	+\frac{1}{U}\,\frac{dU}{d\varphi}\,
	\frac{3\alpha-(2\alpha-1)\varepsilon_1}{2\alpha(2\alpha-1)\varepsilon_1}
	\, \varphi\,^{\prime 2}=0,
	\label{eq800}
\end{equation}
where 
\begin{equation}
	\varepsilon_1=\frac{\alpha}{\ell^2H^2}\left(\frac{H^2  \varphi\,^{\prime 2}}{2}\right)^\alpha ,
	\label{eq802}
\end{equation}
and $H$ is a solution to  
\begin{equation}
	\ell^2	H^2=\frac13\left[	(2\alpha-1)\left(\frac{H^2{\varphi'}^2}{2}\right)^\alpha
	+ U(\lambda\varphi)
	\right].	 
	\label{eq8035}
\end{equation}
Here and from here on, the prime $^\prime$ denotes a derivative with respect to $N$. 

The expression (\ref{eq8035}) is an algebraic equation for the unknown $H^2$ and
is generally not solvable unless $\alpha$
is an integer $<5$.
For a fractional $\alpha$, e.g., $\alpha=1.5$ and $1.3$ used in \cite{papanikolaou}, 
the KG equation involves an implicit function of $H(\varphi,\varphi')$
which requires a cumbersome numerical procedure.
However, one can simplify the numerical procedure by  calculating $H$ via the 2nd Friedmann equation. Then, we need to solve the differential equation 
\begin{equation}
	H'=-\frac{\alpha}{\ell^2 H}\left(\frac{H^2  \varphi\,^{\prime 2}}{2}\right)^\alpha ,
\end{equation}
in addition to 
(\ref{eq800}). 
As a part of this procedure, one needs to solve the algebraic equation (\ref{eq8035})
only at the initial point $N_{\rm in}$ to find the initial value $H_{\rm in}$
as a function of initial $\varphi_{\rm in}$ and $\varphi_{\rm in}'$. 
In any case, we see that
the Hamiltonian approach has a certain advantage,
although the pair of Hamilton equations (\ref{eq7036}) and (\ref{eq7034})
are equivalent to
Eq.\ (\ref{eq800}).

%


\subsection{Model B}

The Lagrangian in this class of  models is of the form
\begin{equation}
	{\cal{L}} = -U(\lambda\varphi)F(X) .
	\label{eq000}
\end{equation}
As in Model A, $\lambda$ is a positive constant of dimension of mass and
$\varphi$ is the field of dimension of length. 
We assume that $F$ is a smooth functions of  
$X\equiv g^{\mu\nu}\varphi_{,\mu}\varphi_{,\nu}>0$, and 
$U$ is an arbitrary smooth dimensionless functions of its argument. 
The corresponding Hamiltonian is given by 
\begin{equation}
	\mathcal{H}=U(\lambda\varphi)\left(F(X)-2XF_X(X)\right).
	\label{eq0041}
\end{equation}
As in Model A, we identify the
Lagrangian and  Hamiltonian with the pressure $p$ and energy density $\rho$, respectively.
Then the Hubble rate is defined by (\ref{eq3060}),
the speed of sound by
\begin{equation}
	c_{\rm s}^2 \equiv 
	\frac{\mathcal{L}_X}{\mathcal{H}_X}
	=\frac{F_X}{F_X+2XF_{XX}},
	\label{eq3029}
\end{equation}
and the first slow-roll parameter by
\begin{equation}
	\varepsilon_1\equiv-\frac{\dot{H}}{H^2}=
	\frac{\mathcal{L}+\mathcal{H}}{2\ell^4 M_{\rm Pl}^2H^2}=
	\frac{3XF_X}{2XF_X-F}.
	\label{eq0046}
\end{equation}

As before, we introduce the conjugate momentum and its magnitude $\eta$.
According to Eq. (\ref{eq6012}),
$X$ is an implicit function of the field $\varphi$ and the conjugate
field $\eta$
\begin{equation}
	4  X(F_X)^2=\frac{\eta^2}{U(\lambda\varphi)^2}.
	\label{eq1012}
\end{equation}
Then, the Hamilton equations (\ref{eq0033})  and (\ref{eq3033})
expressed as differential equations with respect to $N$ in this model are
\begin{equation}
	\frac{d \varphi}{d  N}=-\frac{1}{H}
	\frac{\eta }{2U F_X},
	\label{eq4236}
\end{equation}
\begin{equation}
	\frac{d  \eta }{d  N}= -3\eta   - \frac{F}{H}	\frac{\partial{U}}{\partial\varphi}.
	\label{eq4234}
\end{equation}
It is understood that the variable $X$ in Eqs.\ (\ref{eq4236}), and (\ref{eq4234})
is an implicit function of $\eta^2$ and $\varphi$ via  Eq.\ (\ref{eq1012}). In principle, $X$ may be expressed as an explicit function of $\eta^2$ and $\varphi$ by solving (\ref{eq1012}) for $X$.

The scaling property of the general $k$-essence described by (\ref{eq90077}) applies
directly to this model since we have the freedom to rescale the potential, and rescaling the potential implies identical rescaling of the Lagrangian.
Hence, Model B exhibits invariance of its Hamilton equations
under 
\begin{equation}
	U \rightarrow c_0^{-1}	U, \quad
	\lambda \rightarrow c_0^{-1/2}\lambda, \nonumber
\end{equation}
\begin{equation}
	\varphi \rightarrow c_0^{1/2} \varphi , \quad
	\eta \rightarrow c_0^{-1} \eta .	
	\label{eq9007}
\end{equation}
Clearly, Eq.\ (\ref{eq1012}) renders $X$, $F$, and $F_X$ invariant under this rescaling whereas the Hubble rate scales according to
$H\rightarrow c_0^{-1/2} H$. Then,  the rescaled
Hamilton equations retain the form (\ref{eq4236})
and (\ref{eq4234})
with no dependence on $c_0$.

\subsubsection*{The Tachyon model}
The most popular model that belongs to class B  is the 
Tachyon model
with $F(X)=\sqrt{1-X}$ in Eq.\ (\ref{eq000}) and the Lagrangian 
\begin{equation}
	\mathcal{L}=-U(\lambda\varphi)\sqrt{1-X}.
	\label{eq0038}
\end{equation}
Using this and 
(\ref{eq1012}) we obtain $X$ as an explicit function of the fields $\varphi$ 
and $\eta$ 
\begin{equation}
	X=\frac{\eta^2}{U^2+\eta^2}.
	\label{eq0040}
\end{equation}
The Hamiltonian can be expressed as a function of either $X$ or
$\eta^2$,  
\begin{equation}
	\mathcal{H}= \frac{U}{\sqrt{1-X}}
	=\sqrt{U^2+\eta^2},
\end{equation}
and the Hamilton equations become
\begin{equation}
	\frac{d \varphi}{d  N}=
	\frac{\eta }{H\sqrt{U^2+\eta ^2}} ,
	\label{eq3236}
\end{equation}
\begin{equation}
	\frac{d  \eta }{d  N}= -3\eta   - \frac{U}{H\sqrt{U^2+\eta ^2}}
	\frac{\partial{U}}{\partial\varphi},
	\label{eq3234}
\end{equation}
where the Hubble rate is given by
\begin{equation}
	H^2=
	\frac{\sqrt{U^2+\eta^2}}{3\ell^4 M_{\rm Pl}^2 }.
\end{equation}
The slow-roll parameter and  the sound speed squared can be
expressed as 
\begin{equation}
	\varepsilon_1=\frac32	X= \frac32 \frac{\eta^2}{U^2+\eta^2},
	\label{eq1040}
\end{equation}
\begin{equation}
	c_{\rm s}^2 =1-X=\frac{U^2}{U^2+\eta^2}
	=1-\frac23 \varepsilon_1.
	\label{eq3039}
\end{equation}
%

%

%
%

\section{Curvature perturbations}
\label{perturbations}
We start from  the 1st-order differential equations of Garriga and Mukhanov \cite{garriga} expressed in the form \cite{bilicJCAP,bertini}
\begin{equation}
	a\dot{\xi}_q=z^2c_{\rm s}^2\zeta_q,
	\label{eq3016}
\end{equation}
\begin{equation}
	a\dot{\zeta}_q=-z^{-2}q^2\xi_q,
	\label{eq3017}
\end{equation}
where $q$ is the comoving wavenumber. The quantity $z$ can be expressed in 
different ways:
\begin{equation}
	z^2=\frac{a^2(p+\rho)}{H^2 c_{\rm s}^2} 
	=\frac{2M_{\rm Pl}^2a^2\varepsilon_1}{c_{\rm s}^2} 	
	=\frac{a^2\eta\mathcal{H}_{,\eta}}{\ell^4H^2 c_{\rm s}^2}	
	=\frac{a^2\mathcal{H}_{,\eta}^2}{\ell^4H^2 \mathcal{H}_{,\eta\eta}} .
	\label{eq4029}
\end{equation}
Here, the subscripts $,\!\eta$ and $,\!\eta\eta$ denote respectively the first and second-order partial derivatives with respect to $\eta$.

\subsection{Mukhanov-Sasaki equation}
It is convenient to substitute the e-fold number $N$ for $t$ using $dt=dN/H$ and
express Eqs.\
(\ref{eq3016})-(\ref{eq3017}) as differential equations with respect to
$N$. Then, combining the obtained equations, 
one can derive 
the Mukhanov-Sasaki second-order differential equation 
\begin{equation}
	\frac{d^2 \zeta_q }{dN^2}+\left(3+\varepsilon_2-\varepsilon_1
	-2\frac{c_{\rm s}'}{c_{\rm s}}\right)\frac{d\zeta_q }{dN}
	+\frac{c_{\rm s}^2 q^2}{a^2 H^2}\zeta_q=0,
	\label{eq7001}
\end{equation}
which is equivalent to the set (\ref{eq3016})-(\ref{eq3017}) with 
(\ref{eq4029}). Here, the second slow-roll parameters is defined as $\varepsilon_2=\dot{\varepsilon}_1/(H\varepsilon_1)$.

The perturbations travel at the speed of sound, and their horizon is
the acoustic horizon with radius
$c_{\rm s}/H$.
At the acoustic horizon, the perturbations with the comoving wavenumber $q$ satisfy the horizon crossing relation 
\begin{equation}
	a(N) H(N)= c_{\rm s}(N)q.
	\label{eq50}
\end{equation}
In the slow-roll regime, the perturbations are conserved once they cross the horizon from the subhorizon 
to the superhorizon region. This can be seen as follows.
In the superhorizon region, i.e., 
when the physical wavelength of the perturbation $a/q$ is much bigger than the sound horizon $c_{\rm s}/H$, the last term on the left-hand  side of 
Eq.\ (\ref{eq7001}) can be neglected. Then, in the slow roll approximation, the contribution of the terms proportional to $\varepsilon_1$, $\varepsilon_2$, and
$\dot{c}_{\rm s}$ can also be neglected and the equation takes the form
$\zeta_q''+3\zeta_q'=0$ with solution $\zeta_q=C_1+C_2 \exp (-3N)$, where 
$C_1$ and $C_2$ are constants.
As we shall shortly see, the horizon crossing happens at a relatively large 
$q$-dependent $N$
of the order $N_q\sim 7+\ln q/q_{\rm CMB}$, where $q_{\rm CMB}$ is the CMB pivot scale.
Hence, the curvature perturbations $\zeta_q$ are approximately constant at large $N$ once they cross the horizon 
and enter the superhorizon region where $a(N) H(N)> c_{\rm s}(N)q$, $N>N_q$.

The power spectrum of the curvature perturbations is given by
\begin{equation}
	\mathcal{P}_{\rm S}(q)=\frac{q^3}{2\pi^2}|\zeta_q(N_q)|^2 ,
	\label{eq4024}
\end{equation}
where $\zeta_q(N_q)$ is the solution to Eq.\ (\ref{eq7001})
taken at the point
$N_q$ which for given $q$ satisfy (\ref{eq50}).
The curvature perturbation spectrum  needs to satisfy  
the Harrison-Zeldovich spectrum near $q=q_{\rm CMB}=0.05$ Mpc$^{-1}$
\begin{equation}
	\mathcal{P}_{\rm S}(q)= A_s \left(\frac{q}{q_{\rm CMB}}\right)^{n_s-1},
	\label{eq4025}
\end{equation}
where $A_s=(2.10\pm 0.03)\times 10^{-9}$.
Since the Mukhanov-Sasaki equation (\ref{eq7001}) is linear,
the normalization of $\zeta_q$ can be fixed by the requirement
\begin{equation}
	\mathcal{P}_{\rm S}(q_{\rm CMB})=\frac{q_{\rm CMB}^3}{2\pi^2}|\zeta_{q_{\rm CMB}}(N_{q_{\rm CMB}})|^2=A_s=2.1\cdot 10^{-9}.
	\label{eq5024}
\end{equation}


We need to integrate Eq.\ (\ref{eq7001})
in conjunction with the Hamilton equations in the subhorizon region up to $N_q$,
where $N_q$ satifies the horizon crossing relation $aH=c_{\rm s}q$.
The appropriate $q$-dependent initial point is set in the deep subhorizon region where 
$a_{\rm in}H_{\rm in}<<c_{\rm s}q$.
Hence,  for each wavenumber $q$, we start at a $q$-dependent $N_{{\rm in},q}$ such that
\begin{equation}
	a_{{\rm in},q}H_{{\rm in},q}=\beta c_{{\rm s},q}q,
	\label{eq20}
\end{equation}
where $\beta$ is a small parameter, e.g., $\beta=0.01$ as proposed in Ref.\ \cite{de}.
Here, $H_{{\rm in},q}=H(N_{{\rm in},q})$, $c_{{\rm s},q}=c_{\rm s}(N_{{\rm in},q})$, and
\begin{equation}
	a_{{\rm in},q}=a_0e^{N_{{\rm in},q}-N_0} ,
	\label{eq21}
\end{equation}
where $a_0$ is the cosmological scale factor  at  $N=N_0$.
In particular, for $q=q_{\rm CMB}$, 
\begin{equation}
	a_0e^{N_{{\rm in},q_{\rm CMB}}-N_0}	H_{{\rm in},q_{\rm CMB}}=
	\beta c_{{\rm s},q_{\rm CMB}} q_{\rm CMB}.
	\label{eq36}
\end{equation}

Owing to the $N$-translation invariance of the background equations,
we could choose the origin of inflation at an arbitrary $N_0$
such that the corresponding $q_0<q_{\rm CMB}$, i.e., we set
\begin{equation}
	a_0	H_0=\beta c_{\rm s0}q_0=\epsilon c_{\rm s0}q_{\rm CMB},
	\label{eq32}
\end{equation}
where  
$H_0= H(N_0)$, $c_{\rm s0}= c_{\rm s}(N_0)$,
and the quantity
$\epsilon$ is a small parameter, $\epsilon< \beta$, e.g., $\epsilon=0.001$.
Equation (\ref{eq32}) means that the perturbation with $q=q_{\rm CMB}$ will cross 
the horizon at $N=N_{q_{\rm CMB}}$ which satisfies the condition
\begin{equation}
	\epsilon e^{-N_0}\frac{e^{N_{q_{\rm CMB}}}H(N_{q_{\rm CMB}})}{H_0}\frac{c_{\rm s0}}{c_{\rm s}(N_{q_{\rm CMB}})}=1.
	\label{eq34}
\end{equation}
For an arbitrary $q$, the horizon-crossing relation 
\begin{equation}
	a_0e^{N-N_0} H(N)=c_{\rm s}(N) q,
	\label{eq0021}
\end{equation}
combined with (\ref{eq32}), may be written as
\begin{equation}
	\frac{q}{q_{\rm CMB}}
	=\epsilon	e^{-N_0}\frac{e^N H(N)}{H_0}\frac{c_{\rm s0}}{c_{\rm s}(N)}.
	\label{eq35}
\end{equation}
Hence, the parameter $\epsilon$ always comes in the combination 
$\epsilon	e^{-N_0}$. 
Since $N_0$ is arbitrary, we can conveniently choose,
e.g., $N_0=0$ or a negative number of order $\ln \epsilon$. 
Hence, the dependence on $\epsilon$ may be eliminated by choosing 
a negative $N_0$ of order 10, e.g., $N_0= \ln 0.001=-6.9$.

For $N_0=0$, we have
\begin{equation}
	a_0H_0=\epsilon c_{{\rm s}0}q_{\rm CMB},
	\label{eq22}
\end{equation}
the horizon crossing relation (\ref{eq0021}) becomes 
\begin{equation}
	a_0e^N H(N)=c_{\rm s}(N) q ,
	\label{eq021}
\end{equation} 
and the condition (\ref{eq34}) becomes
\begin{equation}
	\epsilon e^{N_{q_{\rm CMB}}}\frac{H(N_{q_{\rm CMB}})}{H_0}\frac{c_{{\rm s}0}}{c_{\rm s}(N_{q_{\rm CMB}})}=1.
	\label{eq23}
\end{equation}

\subsubsection*{Rough estimates}

During slow-roll inflation, the Hubble rate $H$ and the sound speed $c_{\rm s}$ 
change much slower than the scale factor. In this regime, choosing  $N_{\rm in}=0$ and using (\ref{eq23}), 
we have
a rough estimate for $N_{q_{\rm CMB}}$ at the horizon crossing
\begin{equation}
	N_{q_{\rm CMB}}\simeq \ln \frac{1}{\epsilon} = 6.9.
	\label{eq24}
\end{equation}
Using (\ref{eq20}) and (\ref{eq23}) 
we also find an estimate  for the initial 
$N_{{\rm in},q_{\rm CMB}}$ of the perturbation with $q=q_{\rm CMB}$
\begin{equation}
	N_{{\rm in},q_{\rm CMB}}\simeq \ln \frac{\beta}{\epsilon} 
	= 2.3 .
	\label{eq25}
\end{equation}
Similarly, for any $q\geq q_{\rm CMB}$ we find an order of magnitude estimate  for $N_q$ at the horizon crossing
\begin{equation}
	N_q\simeq \ln \frac{1}{\epsilon}\frac{q}{q_{\rm CMB}}
	= 6.9 +\ln\frac{q}{q_{\rm CMB}} ,
	\label{eq27}
\end{equation}
and for the initial 
$N_{{\rm in},q}$ 
\begin{equation}
	N_{{\rm in},q}	\simeq \ln \frac{\beta}{\epsilon}\frac{q}{q_{\rm CMB}}
	= 2.3 +\ln \frac{q}{q_{\rm CMB}} .
	\label{eq26}
\end{equation}

If we chose $N_{\rm in}=\ln \epsilon =-6.9$, everything would be
shifted by $-6.9$ and we would obtain
\begin{equation}
	N_{q_{\rm CMB}}  \simeq 0,
	\label{eq44}
\end{equation}
\begin{equation}
	N_{{\rm in},q_{\rm CMB}}\simeq \ln \beta 
	= -4.6,
	\label{eq45}
\end{equation}
\begin{equation}
	N_q\simeq 
	\ln\frac{q}{q_{\rm CMB}},
	\label{eq47}
\end{equation}
and
\begin{equation}
	N_{{\rm in},q}	\simeq \ln \beta\frac{q}{q_{\rm CMB}}
	= -4.6  +\ln \frac{q}{q_{\rm CMB}}.
	\label{eq46}
\end{equation}


It is convenient to use 
a new function, sometimes referred to as the Mukhanov-Sasaki variable, $v_q=z\zeta_q$ instead of $\zeta$,
where the quantity $z$ is defined by (\ref{eq4029}).
By making use  Eqs.\ (\ref{eq21}) and (\ref{eq22}),
Eq.\ (\ref{eq7001}) may be written as 
\begin{eqnarray}
	v''_q+(1-\varepsilon_{1})v'_q+\left(
	\frac{e^{-2N}}{\epsilon^2} \frac{c_{\rm s}^2}{c_{{\rm s}0}^2}\frac{H_0^2}{H^2} \frac{q^2}{q_{\rm CMB}^2}
	-(2-\varepsilon_{1}+\frac{\varepsilon_2}{2}-\frac{c_{\rm s}'}{c_{\rm s}})(1+\frac{\varepsilon_2}{2}-\frac{c_{\rm s}'}{c_{\rm s}})
	\right.
	\nonumber \\
	\left.		
	-\frac{\varepsilon_{2}\varepsilon_3}{2}+(\frac{c_{\rm s}'}{c_{\rm s}})'\right)v_q=0,
	\label{eq7002}
\end{eqnarray}
where the prime $^\prime$ denotes a derivative with respect to $N$ and should not be confused with the derivative with respect to the conformal time. The slow roll parameters are defined as $\varepsilon_{1}\equiv -\dot{H}/H^2=-H'/H$ and $\varepsilon_{i+1}\equiv \dot{\varepsilon}_{i}/(H\varepsilon_{i})=\varepsilon_{i}^{\prime}/\varepsilon_{i}$ for $i>1$.

To find the spectrum at the horizon crossing, equation (\ref{eq7002}) should  be integrated for a fixed $q$ up to the e-fold number
$N_{\rm f}$ which satisfies the horizon crossing equation 
(\ref{eq021}). If we fix the initial point at $N_{\rm in}=0$ and combine  Eqs.\
(\ref{eq22}) and (\ref{eq021})
we obtain 
\begin{equation}
	N= \ln \frac{1}{\epsilon} \frac{q}{q_{\rm CMB}} 
	\frac{H_0}{H(N)}\frac{c_{\rm s}(N)}{c_{{\rm s}0}},
	\label{eq7006}
\end{equation}
which, given $q$, may be regarded as an algebraic equation for $N$. 
The explicit 
functional dependence $H(N)$ and $c_{\rm s}(N)$
in (\ref{eq7006}) is obtained by integrating the background Hamilton equations
in parallel with (\ref{eq7002}).
Then, the spectrum $\mathcal{P}_{\rm S}(q)$ can be plotted using the solutions $\zeta_q(N_{\rm f})$ at the point $N_{\rm f}$ that satisfies (\ref{eq7006})
for each $q$.

More precisely, the calculation of the spectrum proceeds as follows.  For a set of fixed $N_{\rm in}$,
e.g, $N_{\rm in}=0,1,2,3, ...$,  integrate 
the background equations up to $N_{\rm in}$ to find $H(N_{\rm in})/H_0$ and 
$c_{\rm s}(N_{\rm in})/c_{{\rm s}0}$. Using these, calculate 
$q/q_{\rm CMB}$ for each $N_{\rm in}$ from the equation
\begin{equation}
	\frac{q}{q_{\rm CMB}}= \frac{\epsilon}{\beta} 
	\frac{H(N_{\rm in})}{H_0}\frac{c_{{\rm s}0}}{c_{\rm s}(N_{\rm in})}e^{N_{\rm in}},
	\label{eq7007}
\end{equation}
obtained by combining Eqs.\ (\ref{eq20}), (\ref{eq21}), and (\ref{eq32}).
Then, solve
(\ref{eq7006}) for $N$  to find the horizon crossing point $N_{\rm f}$ for each $q$. 
Finally,  integrate  (\ref{eq7002}) from $N_{\rm in}$ to $N_{\rm f}$ together with the Hamilton equations and
plot $\mathcal{P}_{\rm S}$ versus $q/q_{\rm CMB}$ using $\zeta_q(N_{\rm f})$
for each $q/q_{\rm CMB}$.

\subsubsection*{Initial conditions}

To determine the proper initial conditions in the deep subhorizon region, we adopt the standard Bunch-Davies vacuum solution \cite{bunch}  
\begin{equation}
	v_q(\tau)=\frac{e^{-ic_{\rm s}q\tau}}{\sqrt{2c_{\rm s}q}} ,
	\label{eq1}
\end{equation} 
where the conformal time $\tau=\int dt/a$ in the slow-roll regime satisfies
\begin{equation}
	\tau=-\frac{1+\varepsilon_1}{aH}+\mathcal{O}(\varepsilon_1^2) ,
	\label{eq2}
\end{equation}
with $-qc_{\rm s}\tau \gg 1$ in the deep subhorizon region. 
Hence, the initial values of $v_q$ and $v_q'$ at $N_{\rm in}$  are determined 
by (\ref{eq1}) up to an arbitrary phase.  The simplest choice is 
\begin{equation}
	v_{q\rm in}\equiv v_q(N_{\rm in})=\frac{1}{\sqrt{2c_{\rm s}q}},
	\label{eq10}
\end{equation}
\begin{equation}
	v'_{q\rm in}\equiv v_q'(N_{\rm in})=-i\frac{c_{\rm s}q}{a_{\rm in}H_{\rm in}\sqrt{2c_{\rm s}q}}.
	\label{eq11}
\end{equation}
Then, according to (\ref{eq20}), we set
\begin{equation}
	a_{\rm in}H_{\rm in}=\beta qc_{\rm s},
\end{equation}
so 
\begin{equation}
	v'_{q\rm in}=-i\frac{1}{\beta\sqrt{2c_{\rm s}q}}.
	\label{eq13}
\end{equation}
It is understood that the function $c_{\rm s}$ 
in Eqs.\ (\ref{eq10})-(\ref{eq13}) 
are taken at $N=N_{\rm in}$.

\subsection{Redefinition of the input parameters and fields}
\label{renormalize}

In our calculations, we assume certain values of the parameters and initial values of the fields. Besides, a solution to Eq.\ (\ref{eq7002}) involves
the Bunch-Davies initial conditions (\ref{eq10}) and (\ref{eq11}). These initial conditions determine the normalization of 
$\zeta_q(N)$ which in turn fixes the normalization of $\mathcal{P}_{\rm S}$. 
Hence,
there is {\it a priori} no guarantee that the obtained spectrum will satisfy 
the condition (\ref{eq5024}). 
Instead, the calculated value of the spectrum   $\mathcal{P}_{\rm S}(q)$ at the point
$q=q_{\rm CMB}$ will satisfy
\begin{equation}
	\mathcal{P}_{\rm S}(q_{\rm CMB})=\frac{A_s}{c_0},
\end{equation}
where $c_0$ is a constant, generally $c_0\neq 1$. 
Thus, we have a conflict between the imposed Bunch-Davies vacuum and 
the normalization of the spectrum to the observed value at the CMB pivot scale.  

To rectify this conflict, 
we need an appropriate redefinition of the input parameters 
by making use 
of the scaling invariance of the background equations.
To begin, 
we multiply  $\mathcal{P}_{\rm S}$ by $c_0$ and write
\begin{equation}
	\mathcal{P}_{\rm S}=c_0\frac{q^3}{2\pi^2}|\zeta_q(N_q)|^2 = 
	\frac{c_0 H^2}{4\pi^2\varepsilon_1 M_{\rm Pl}^2}
	q|v_q|^2,
	\label{eq66}
\end{equation}
where we have used Eqs.\ (\ref{eq4029}) and  (\ref{eq50}).
Now we absorb $c_0$ in $H^2$ 
and rescale the Hubble rate as $H\rightarrow c_0^{-1/2}  H$.
Note that this rescaling does not affect Eq.\ (\ref{eq7002}) and its solution $v_q$
because each term in (\ref{eq7002}) is invariant under multiplying $H$ by a constant. This invariance can 
easily be verified by checking term by term. The term containing $H/H_0$ is invariant since $H_0$ scales in the same way as $H$.  Likewise, the quantities $\varepsilon_1\equiv -H’/H$, $\varepsilon_2\equiv \varepsilon'_1/\varepsilon_1$, and
$\varepsilon_3\equiv \varepsilon'_2/\varepsilon_2$  are also invariant under multiplying $H$ by a constant. Finally, the sound speed and its derivatives are not affected  due to (\ref{eq3019}) with (\ref{eq3060}).

Thus, we obtain a properly normalized spectrum without the $c_0$ factor. However, rescaling $H$, we need to redefine the model's input parameters and background field initial values. 
Using  the scale invariance (\ref{eq8007}) (Model A) or (\ref{eq9007}) (Model B) of the background  
we infer a redefinition of the parameters.  
If we repeat the calculations using the redefined input parameters
$V_0$, $\lambda$, and initial values rescaled according to (\ref{eq8007})
in Model A or  (\ref{eq9007}) in Model B, we will obtain the power spectrum that agrees with the observed value at the CMB pivot scale.

\subsection{Approximate spectrum}
\label{approximate}

In the slow-roll regime, the curvature spectrum can be approximated by
(see, e.g., Ref.\ \cite{bertini})
\begin{equation}
	\mathcal{P}_{\rm S}(q)\simeq \frac{1}{8\pi^2c_{\rm s}\varepsilon_1}
	\frac{H^2}{M_{\rm Pl}^2},
	\label{eq6021}
\end{equation}
where for each $q$, the quantities $H$, $ c_{\rm s}$, and $\varepsilon_1$ take on their horizon crossing values at the corresponding $N$. 
The approximate spectrum obtained in this way will not in general
have the correct observed value at the pivot scale $q_{\rm CMB}$.
To satisfy the proper normalization of $\mathcal{P}_{\rm S}(q)$ at $q=q_{\rm CMB}$,  we can 
introduce a constant factor $\bar{c}_0$. To wit, we define
\begin{equation}
	\mathcal{P}_{\rm S}(q)\simeq \frac{\bar{c}_0}{8\pi^2c_{\rm s}\varepsilon_1}
	\frac{H^2}{M_{\rm Pl}^2},
	\label{eq5021}
\end{equation}
and fix $\bar{c}_0$ so that $\mathcal{P}_{\rm S}(q_{\rm CMB})$ has the correct 
observed value at the CMB pivot scale.
However, a choice  $\bar{c}_0\neq 1$ will generally violate the assumed Bunch-Davies asymptotic behavior. 
As before, this problem can be sorted out
by absorbing $\bar{c}_0$ in $H^2$ and redefining the physical parameters in such a way that the equation for the approximate spectrum reads
as in (\ref{eq6021}).
Repeating the procedure of section \ref{renormalize}, we redefine the parameters and fields as in Eqs.\ (\ref{eq8007}) (Model A) or  Eqs.\ (\ref{eq9007}) (Model B), in which $c_0$ is replaced by 
$\bar{c}_0$.

The constants $c_0$ and $\bar{c}_0$ need not be necessarily equal.
However, if we want to compare the exact with the approximate spectrum, we have to stick to the same parameterization of the model in both exact and approximate calculations. In this case, we will replace $\bar{c}_0$ in (\ref{eq5021}) by $c_0$ so that the redefinition of parameters and initial values will be the same.
Obviously, in this case, the approximate spectrum will fail to reproduce the correct
observed value at $q=q_{\rm CMB}$.

\subsection{Integration of the Mukhanov-Sasaki equation}
To integrate Eq.\ (\ref{eq7001})
in conjunction with the Hamilton equations
we need to choose a specific model.
By way of example, we present here the spectra calculated for two specific
models representing respectively classes A and B. 
Class A will be represented by the PLLS model of Ref. \cite{papanikolaou} with
the Lagrangian (\ref{eq0035}) in which $\alpha=1.5$. 
Class B will be represented by the tachyon Lagrangian (\ref{eq0038}).
In both models, we will use 
a potential with inflection point proposed in Ref.\ \cite{papanikolaou} 
\begin{equation}
	U(\lambda\varphi)=V(\lambda\varphi)-V(0),
	\label{eq1135}
\end{equation}
where
\begin{equation}
	\quad
	V(\lambda\varphi)=\ell^4 V_0\exp\left(-\frac{|\kappa-\lambda\varphi|^n {\text{sgn}}(\kappa-\lambda\varphi)}{\ell^{n}M_{\rm Pl}^n}\right). 
	\label{eq1035}
\end{equation}
The quantity $V_0$ is a constant of dimension of mass to the 4th power.
The dimensionless constant $\kappa$ corresponds to a shift $\phi_0=\lambda^{-1}\ell^{-2}\kappa$ of the physical fiel $\phi$. 
 
Owing to the inflection point of the potential at $\varphi=\kappa/\lambda$, the scalar field will exhibit an extended flat region for about 30 e-folds. In this flat region, slow-roll conditions will not hold, and inflation will enter into a temporary ultra-slow-roll regime. During this period, the non-constant mode of the curvature fluctuations, which would decay exponentially in the slow-roll regime, actually grows in the ultra-slow-roll regime, enhancing in this way the curvature power spectrum at specific scales that can potentially collapse, forming primordial BHs in the early Universe (for details, see Ref. \cite{papanikolaou}).

\subsubsection*{Input parameters}
The initial curvature perturbations will eventually collapse to form BHs if the peak in the amplitude exceeds a certain threshold that strongly depends on the spectrum shape \cite{germani}. 
A desirable spectrum requires the tuning of
the input parameters and initial values.
To this end,
we  will use the parameterization  of Ref.\ \cite{papanikolaou}
and 
fix $\ell=10^6 M_{\rm Pl}^{-1}$, $n=3$, 
$V_0=10^{-16}M_{\rm Pl}^4$, and $\lambda
=1.961\cdot 10^{-8} M_{\rm Pl}$.
Here, we have a slight notation difference compared to Ref.\ \cite{papanikolaou}. 
First, the mass scale $M$ in \cite{papanikolaou} corresponds
to  $\ell^{-1}$ in our notation.
Second, 
the dimensionless quantity $(\lambda\ell)^n$ is identical to the quantity denoted by $\lambda$ in \cite{papanikolaou}.
Thus, the above chosen $\lambda =1.961\cdot 10^{-8} M_{\rm Pl}$   corresponds to $\lambda=7.54\cdot 10^{-6}$ of Ref.\ \cite{papanikolaou}.
Third, the constant shift $\phi_0$  of the physical field $\phi$ is identical to our $\lambda^{-1} \ell^{-2}\kappa$.  
The remaining  input parameters are  listed in Table \ref{table1} for each model
in the columns entitled ``original". 

\begin{figure}[h!]
	\begin{center}
		\includegraphics[scale=0.25]{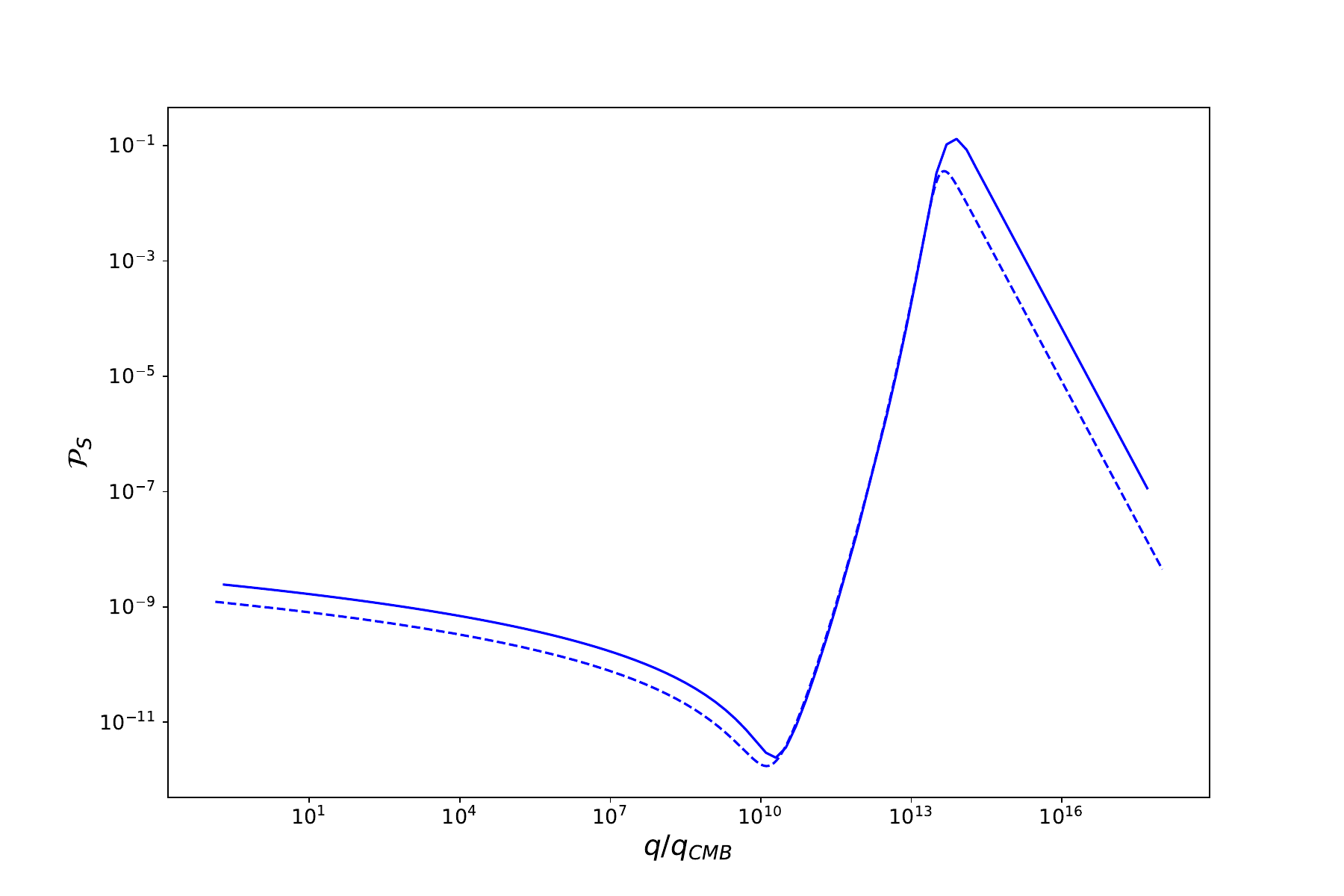}
		\includegraphics[scale=0.25]{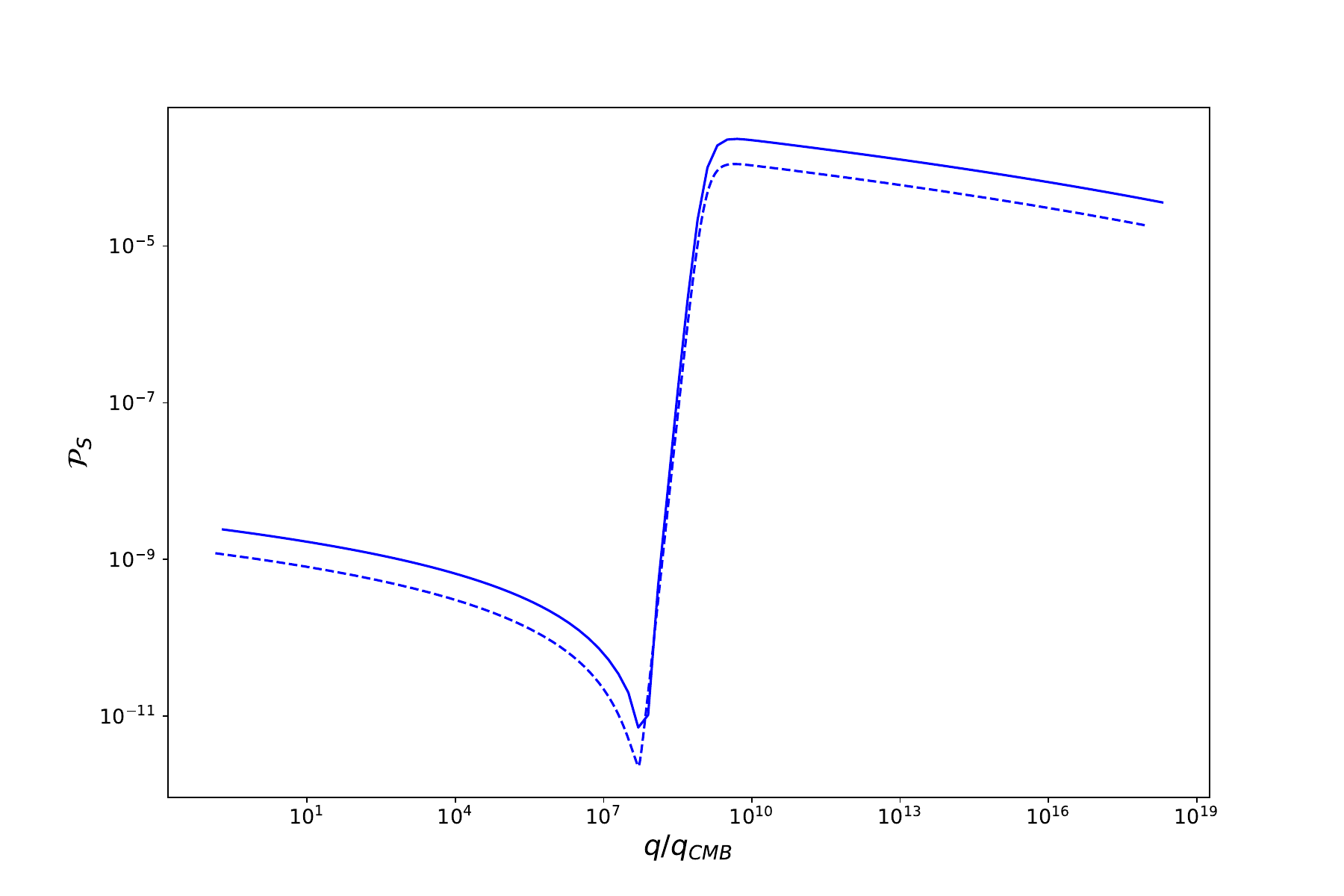}
		\caption{
			The curvature power spectrum obtained by
			numerically solving Eq.\ (\ref{eq7002}) 
			(full line) combined with the  spectrum  approximated by 
			Eq.\ (\ref{eq5021}) (dashed line) for the PLLS model (left panel) and 
			Tachyon model (right panel). 
			The input parameters are provided in Table \ref{table1} for each model.
		}
		\label{fig1}
	\end{center}
\end{figure}

\begin{figure}[h!]
	\begin{center}
		\includegraphics[scale=0.8]{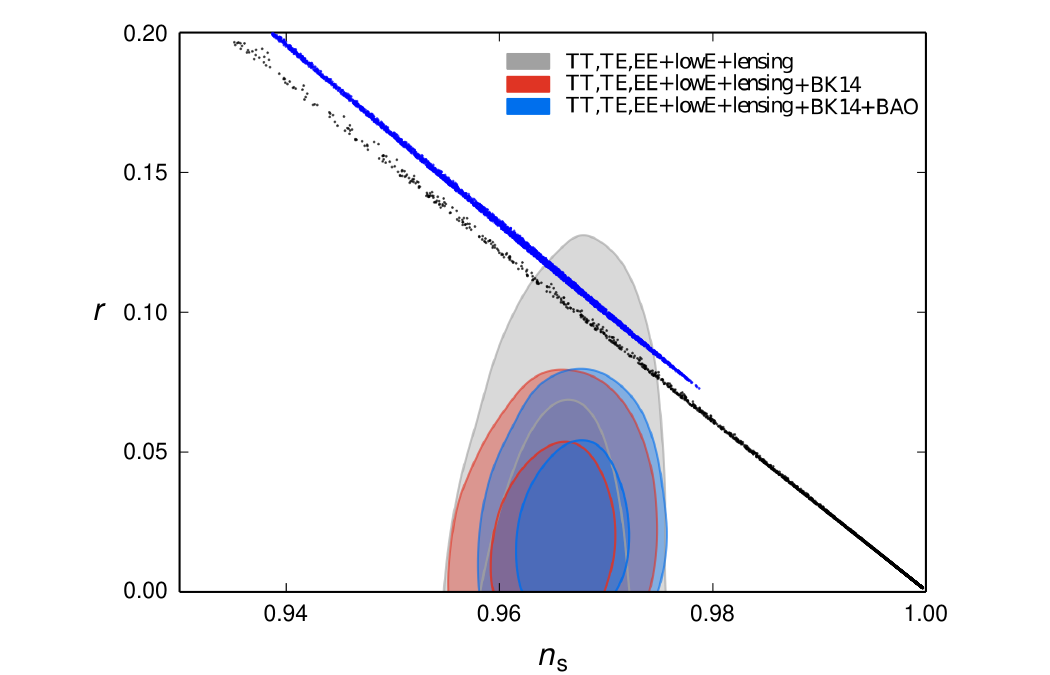}
		\caption{
			$r$ versus $n_{\rm{S}}$ diagram with observational constraints from Ref.\ \cite{Planck:2018jri}. The dots represent
			the theoretical predictions of the PLLS model (blue) and Tachyom model (black).
			The dots are obtained by varying the parameters $\phi_0$ and $\lambda$ of the potential  and initial values $\phi_{\rm in}$ and  $\eta_{\rm in}$ as discussed in the text.
				}
		\label{fig2}
	\end{center}
\end{figure}

Using these parameters, we find that the proper normalization
to the CMB pivot value at $q=q_{\rm CMB}$, as discussed before, 
requires $c_0=9.4732\cdot 10^{10}$ for the PLLS model and $c_0= 10.972$ for the Tachyon model.
Hence, to obtain a correct model parameterization, we need to appropriately
redefine the parameters and initial values according to (\ref{eq8007}) for the PLLS model and (\ref{eq9007}) for the Tachyon model. 
The redefined values are provided in Table \ref{table1} for each model
in the columns entitled ``redefined".

\begin{table}
\caption{Input parameters}
 \centering
\begin{tabular}{||l|c|c|c|c||} \hline
	&     \multicolumn{2}{c|}{PLLS model, $\alpha=1.5$} & \multicolumn{2}{c||}{Tachyon model}  \\ \hline
	& original &redefined &  original &redefined \\ \hline   
$V_0 [M_{\rm Pl}^4]$ & $10^{-16}$  & $9.4732\cdot 10^{-6}$& 
$10^{-16}$ & $1.0972\cdot 10^{-15} $  \\ \hline
$ \lambda [M_{\rm Pl}]$ & $1.961\cdot 10^{-8}$ & $1.324\cdot 10^{-6}$ &
$1.961\cdot 10^{-8}$ & $6.49561\cdot 10^{-8}$ 
\\ \hline 
$ \phi_0 [M_{\rm Pl}]$ & $0.835$ & $0.0124$ &
$0.2586$ & $0.0781$
\\ \hline 
$ \phi_{\rm in} [M_{\rm Pl}]$ & $5.3$ & $0.0785$ &
$1.1$ & $0.3321$
\\ \hline
$\eta_{\rm in}$ & $-1.53\cdot 10^{-8}$ & $-0.3179$ &
$-3\cdot 10^{-5}$ & $-3.3\cdot 10^{-4}$
\\ \hline
\end{tabular}
\label{table1}
\end{table}

In this way, if we repeat the calculations using the redefined input parameters
$V_0$, $\lambda$, and initial values,  we will automatically obtain the 
properly normalized power spectra
 with no need for the normalization constant $c_0$.
In Fig.\ \ref{fig1}, we plot the curvature power spectra obtained by
numerically solving equation (\ref{eq7002}) 
combined with the  spectra  approximated by 
Eq.\ (\ref{eq5021})  for the PLLS and Tachyon  models, both  with the same potential defined 
by  (\ref{eq1135}) and (\ref{eq1035}) and redefined input parameters
provided in Table \ref{table1}.

Although we are mainly concerned with the power spectrum normalization, it is of considerable interest to explicitly compare the predicted scalar spectral index $n_{\rm S}$ and the tensor-to-scalar ratio $r$ of each model with current observational constraints.
To this end, we compute $n_{\rm S}$ and $r$ for each model for different sets of model parameters and initial values.
Since we have not calculated the tensor perturbation spectra,
we compute  using  the linear approximation in $\varepsilon$-parameters \cite{Lola:2020lvk}   
\begin{eqnarray}
	n_{\rm S} & \approx & 1 - 2\varepsilon_1 - \varepsilon_2,\\
	r & \approx & 16 c_s \varepsilon_1,
\end{eqnarray}
where $\varepsilon_1$ and $\varepsilon_2$ are evaluated
at the acoustic horizon of the perturbations with the comoving wavenumber $q_{\rm CMB}$.
We have randomly generated many sets of input parameters and initial values in the intervals 
$0.2 < \phi_0 [M_{\rm Pl}]< 0.3$, $10^{-8} < \lambda [M_{\rm Pl}]< 9.9\cdot10^{-8}$,
$1.0 < \phi_{\rm in}[M_{\rm Pl}] < 1.2$, and $-10^{-4} < \eta_{\rm in} < -9.9\cdot 10^{-6}$  for the Tachyon model, and $0.75 < \phi_0 [M_{\rm Pl}]< 0.9$, , $2.7\cdot 10^{-8} < \lambda[M_{\rm Pl}] < 9.9\cdot 10^{-8}$, $4.5 < \phi_{\rm in} [M_{\rm Pl}]< 5.5$, and $-10^{-6} < \eta_{\rm in} < -9.9\cdot 10^{-9}$ for the PLLS model. The parameters $V_0$ and $\alpha$ given in Table~\ref{table1} are kept fixed. The resulting value sets for $n_{\rm S}$ and $r$ are collected in the plot in Fig.~\ref{fig2} together with observational constraints from Planck 2018 combined data \cite{Planck:2018jri}.

The numerical calculations and plots of the power spectrum reveal 
that the approximate solution is consistently lower by a factor of 3 - 5. The most discrepancy occurs at the minimum of $P_{\rm S}$.
Despite the overall disagreement, the approximate solution 
captures the essential shape and features of the numerical solution.
The disagreement would be much lower if the approximate spectrum were properly normalized to fit the correct value  
at the CMB pivot scale. In this case, as discussed in Sec.\ \ref{approximate}, the normalization constant $\bar{c}_0$ would differ from $c_0$. Consequently, the redefined parameterization would not reproduce the same original model.
The main advantage of the approximate solution is that it considerably 
simplifies and accelerates numerical calculations.
 
The curvature power spectrum could be fine-tuned to achieve a desirable PBH formation as in single-field inflationary models \cite{Cole:2023wyx}. Besides, fine-tuning might be needed to improve the model predictions of observational parameters, such as the spectral index $n_{\rm S}$ and	scalar-to-tensor ratio $r$. Although, in this work, we have not been concerned with this fine-tuning issue, we have checked the sensitivity of the spectra to the input parameters and initial conditions by varying the parameters and initial conditions for the background up to 20\%. These variations resulted in transparent and expected changes in the calculated power spectra. Given the initial conditions, the resulting power spectra in both models are smooth and consistent with the approximated power spectrum. As shown in Fig.\ \ref{fig1}, the spectra have transparent shapes with a distinct peak. When the initial conditions are varied by up to 20\%, the spectra exhibit a visible change, particularly at the peak of the power spectrum. The overall shape remains almost unchanged, but the peak's magnitude and position are noticeably affected.

It is important to mention that the backreaction of small-scale one-loop corrections to the large-scale curvature power spectrum could potentially alter the curvature perturbation amplitude measured by Planck.  A careful analysis of this backreaction issue,  extensively studied in the literature \cite{Inomata:2022yte,Kristiano:2022maq,Choudhury:2023rks,Ballesteros:2024zdp,Franciolini:2023agm,Firouzjahi:2023ahg}, would go beyond the scope of our paper. At present, we can only say that one-loop corrections could significantly alter the curvature perturbations in the models studied here. However, since these corrections do not affect the background,  the scaling properties still hold and, we believe, the input parameters and initial values could be redefined to maintain the correct amplitude at the CMB pivot scale.

\subsubsection*{Note on the numerical routines}

We have conducted thorough tests of different numerical integration methods for solving the Hamilton equations and Mukhanov-Sasaki equation, specifically comparing the results from the Runge-Kutta-Fehlberg 7-8, Runge-Kutta Cash-Karp 5-4, and the classic 4th-order Runge-Kutta (RK4) methods. These tests revealed that the average relative differences among these methods' solutions are approximately on the order of $10^{-5}$, demonstrating low sensitivity to the choice of the numerical integrator. In all final calculations, we have used the Runge-Kutta-Fehlberg 7-8 method.

\section{Summary and conclusions}
\label{conclude}

We have demonstrated that a general $k$-essence exhibits a simple  scaling property
provided the Lagrangian admits a multiplication by a constant.
In particular, we have studied the curvature perturbation spectra for two broad classes of $k$-inflation models.
We have demonstrated that both A and B models enjoy scaling properties, which
one can use to tune the model parameters and initial values 
without affecting the shape of the perturbation spectrum.
However, a tuning of the input parameters, often necessary to achieve the desired properties of the spectrum, will inevitably change the spectrum normalization
if one adheres to the Bunch-Davies asymptotic condition.

Using specific representatives of the A and B models, we have shown that our initial choice of input parameters 
has led to a large disagreement with
the observed value  at the CMB pivot scale
by a factor of the order $10^{11}$ in the PLLS model and $10$ in the Tachyon model. 
However, owing to the scaling properties, we are allowed to redefine the input parameters 
in such a way that the spectra keep their shape, satisfy the Bunch-Davies asymptotic, 
and agree with the observational value at the CMB pivot scale. 

We have calculated the scalar spectral index ($n_{\rm S}$) and tensor-to-scalar ($r$) ratio 
for a large number of sets of input parameters in the two models and presented a comparison with Planck 2018 data. 
Besides, we have checked the sensitivity of the curvature power spectrum to the variation of 
the input parameters and initial conditions.  

Our analysis may be useful in the search for inflation models in the context of primordial black-hole production.

\subsection*{Acknowledgments}
This work, in its initial phase, was supported by the ICTP – SEENET-MTP project NT-03 "Theoretical and Computational Methods in Gravitation and Astrophysics – TECOM-GRASP" and the COST Action CA1810, "Quantum gravity phenomenology in the multi-messenger approach".  M.\ Stojanovi\'{c} acknowledges the support by the Ministry of Science, Technological Development and Innovation of the Republic of Serbia under contract 451-03-137/2025-03-200113. D.D.\ Dimitrijevi\'{c}, G.S.\ Djordjevi\'{c}, and M.\ Milo\v{s}evi\'{c} acknowledge the support by the same Ministry under contract 451-03-137/2025-03/200124.  G.S.\ Djordjevi\'c and D.D.\ Dimitrijevi\' c acknowledge the support by the CEEPUS Program RS-1514-05-2425 "Quantum Spacetime, Gravitation and Cosmology". G.S.\ Djordjevi\'c acknowledges the hospitality of the CERN-TH. N.\ Bili\'c acknowledges the hospitality of the SEENET-MTP Centre and the Department of Physics, University of Ni\v{s}, where part of his work was completed.


\begin{thebibliography}{90}
	
	
\bibitem{armendariz2}
C.~Armendariz-Picon, V.~F.~Mukhanov and P.~J.~Steinhardt,
Phys.\ Rev.\ Lett.\  {\bf 85} (2000) 4438
[arXiv:astro-ph/0004134].	

\bibitem{armendariz1}
C.~Armendariz-Picon, T.~Damour and V.~F.~Mukhanov,
Phys.\ Lett.\ B {\bf 458} (1999) 209
[arXiv:hep-th/9904075].

\bibitem{kamenshchik}
A.~Y.~Kamenshchik, U.~Moschella and V.~Pasquier,
Phys. Lett. B \textbf{511}, 265-268 (2001)
[arXiv:gr-qc/0103004 [gr-qc]].

\bibitem{bilic}
N.~Bilic, G.~B.~Tupper and R.~D.~Viollier,
Phys. Lett. B \textbf{535}, 17-21 (2002)
[arXiv:astro-ph/0111325 [astro-ph]].

\bibitem{fairbairn} 
M.~Fairbairn and M.~H.~G.~Tytgat,
{\em Inflation from a tachyon fluid?},
Phys.\ Lett.\ B {\bf 546}, 1 (2002)
[hep-th/0204070].

\bibitem{frolov} 
A.~V.~Frolov, L.~Kofman and A.~A.~Starobinsky,
{\em Prospects and problems of tachyon matter cosmology},
Phys.\ Lett.\ B {\bf 545}, 8 (2002)
[hep-th/0204187].

\bibitem{shiu1} 
G.~Shiu and I.~Wasserman,
{\em Cosmological constraints on tachyon matter},
Phys.\ Lett.\ B {\bf 541}, 6 (2002)
[hep-th/0205003]. 

\bibitem{sami} 
M.~Sami, P.~Chingangbam and T.~Qureshi,
{\em Aspects of tachyonic inflation with exponential potential},
Phys.\ Rev.\ D {\bf 66}, 043530 (2002)
[hep-th/0205179];
P.~Chingangbam, S.~Panda and A.~Deshamukhya,
{\em Non-minimally coupled tachyonic inflation in warped string background},
JHEP {\bf 0502}, 052 (2005)
[hep-th/0411210].

\bibitem{shiu2} 
G.~Shiu, S.~H.~H.~Tye and I.~Wasserman,
{\em Rolling tachyon in brane world cosmology from superstring field theory},
Phys.\ Rev.\ D {\bf 67}, 083517 (2003)
[hep-th/0207119];
S.~del Campo, R.~Herrera and A.~Toloza,
{\em Tachyon Field in Intermediate Inflation},
Phys.\ Rev.\ D {\bf 79}, 083507 (2009)
[arXiv:0904.1032];
S.~Li and A.~R.~Liddle,
{\em Observational constraints on tachyon and DBI inflation},
JCAP {\bf 1403}, 044 (2014)
[arXiv:1311.4664]. 

\bibitem{kofman} 
L.~Kofman and A.~D.~Linde,
{\em Problems with tachyon inflation},
JHEP {\bf 0207}, 004 (2002)
[hep-th/0205121].

\bibitem{cline} 
J.~M.~Cline, H.~Firouzjahi and P.~Martineau,
{\em Reheating from tachyon condensation},
JHEP {\bf 0211}, 041 (2002)
[hep-th/0207156].

\bibitem{salamate2018}
F.~Salamate, I.~Khay, A.~Safsafi, H.~Chakir and M.~Bennai,
{\em Observational Constraints on the Chaplygin Gas with Inverse Power Law Potential in Braneworld Inflation}, 
Mosc. Univ. Phys. Bull. {\bf 73} 405 (2018).

\bibitem{barbosa2018}
N.~Barbosa-Cendejas, R.~Cartas-Fuentevilla, A.~Herrera-Aguilar, R.~R.~Mora-Luna and R.~da Rocha,
{\em A de Sitter tachyonic braneworld revisited},
JCAP. {\bf 2018} 005 (2018)
[hep-th/1709.09016].

\bibitem{dantas2018}
D.~M.~Dantas, R.~da Rocha and C.~A.~S.~Almeida,
{\em Monopoles on string-like models and the Coulomb's law},
Phys. Lett. B. {\bf 782} 149 (2018)
[hep-th/1802.05638].

\bibitem{steer}
D.~A.~Steer and F.~Vernizzi,
{\em Tachyon inflation: Tests and comparison with single scalar field inflation},
Phys.\ Rev.\ D {\bf 70}, 043527 (2004)
[hep-th/0310139].

\bibitem{bilicJCAP}
N.~Bilic, D.~D.~Dimitrijevic, G.~S.~Djordjevic, M.~Milosevic and M.~Stojanovic,
JCAP \textbf{08}, 034 (2019)
[arXiv:1809.07216 [gr-qc]].

\bibitem{papanikolaou}
T.~Papanikolaou, A.~Lymperis, S.~Lola and E.~N.~Saridakis,
JCAP \textbf{03} (2023), 003
[arXiv:2211.14900 [astro-ph.CO]].

\bibitem{garcia-bellido}
J.~Garcia-Bellido and E.~Ruiz Morales,
Phys. Dark Univ. \textbf{18}, 47-54 (2017)
[arXiv:1702.03901 [astro-ph.CO]].

\bibitem{kannike}
K.~Kannike, L.~Marzola, M.~Raidal and H.~Veerm\"ae,
JCAP \textbf{09}, 020 (2017)
[arXiv:1705.06225 [astro-ph.CO]].

\bibitem{germani}
C.~Germani and I.~Musco,
Phys. Rev. Lett. \textbf{122}, no.14, 141302 (2019)
[arXiv:1805.04087 [astro-ph.CO]].

\bibitem{gibbons} 
G.~W.~Gibbons,
{\em Thoughts on tachyon cosmology},
Class.\ Quant.\ Grav.\  {\bf 20}, S321 (2003)
[hep-th/0301117].

\bibitem{sen} 
A.~Sen,
{\em Supersymmetric world volume action for nonBPS D-branes},
JHEP {\bf 9910}, 008 (1999)
[hep-th/9909062].

\bibitem{Sen:2002in}
A.~Sen,
JHEP \textbf{07} (2002), 065
doi:10.1088/1126-6708/2002/07/065
[arXiv:hep-th/0203265 [hep-th]].

\bibitem{shandera} 
S.~E.~Shandera and S.-H.~H.~Tye,
JCAP {\bf 0605}, 007 (2006)
[hep-th/0601099].

\bibitem{bilic2}
N.~Bili\'c, S.~Domazet and G.~Djordjevic,
Class. Quant. Grav. \textbf{34}, no.16, 165006 (2017)
[arXiv:1704.01072 [gr-qc]].

\bibitem{garriga}
J.~Garriga and V.~F.~Mukhanov,
Phys. Lett. B \textbf{458}, 219-225 (1999)
doi:10.1016/S0370-2693(99)00602-4
[arXiv:hep-th/9904176 [hep-th]].

\bibitem{bertini}
N.~R.~Bertini, N.~Bilic and D.~C.~Rodrigues,
Phys. Rev. D \textbf{102}, no.12, 123505 (2020)
[erratum: Phys. Rev. D \textbf{105}, no.12, 129901 (2022)]
[arXiv:2007.02332 [gr-qc]].

\bibitem{de}
A.~De and R.~Mahbub,
Phys. Rev. D \textbf{102}, no.12, 123509 (2020)
[arXiv:2010.12685 [astro-ph.CO]].

\bibitem{bunch}
T.~S.~Bunch and P.~C.~W.~Davies,
Proc. Roy. Soc. Lond. A \textbf{360}, 117-134 (1978).

\bibitem{Lola:2020lvk}
S.~Lola, A.~Lymperis and E.~N.~Saridakis,
Eur. Phys. J. C \textbf{81}, no.8, 719 (2021) 
[arXiv:2005.14069 [gr-qc]].	

\bibitem{Planck:2018jri}
Y.~Akrami \textit{et al.} [Planck],
Astron. Astrophys. \textbf{641}, A10 (2020)
[arXiv:1807.06211 [astro-ph.CO]].

\bibitem{Cole:2023wyx}
P.~S.~Cole, A.~D.~Gow, C.~T.~Byrnes and S.~P.~Patil,
JCAP \textbf{08}, 031 (2023)
[arXiv:2304.01997 [astro-ph.CO]].

\bibitem{Inomata:2022yte}
K.~Inomata, M.~Braglia, X.~Chen and S.~Renaux-Petel,
JCAP \textbf{04}, 011 (2023)
[erratum: JCAP \textbf{09}, E01 (2023)]
doi:10.1088/1475-7516/2023/04/011
[arXiv:2211.02586 [astro-ph.CO]].

\bibitem{Kristiano:2022maq}
J.~Kristiano and J.~Yokoyama,
Phys. Rev. Lett. \textbf{132}, no.22, 221003 (2024)
doi:10.1103/PhysRevLett.132.221003
[arXiv:2211.03395 [hep-th]].

\bibitem{Choudhury:2023rks}
S.~Choudhury, S.~Panda and M.~Sami,
JCAP \textbf{11}, 066 (2023)
doi:10.1088/1475-7516/2023/11/066
[arXiv:2303.06066 [astro-ph.CO]].

\bibitem{Ballesteros:2024zdp}
G.~Ballesteros and J.~G.~Egea,
JCAP \textbf{07}, 052 (2024)
doi:10.1088/1475-7516/2024/07/052
[arXiv:2404.07196 [astro-ph.CO]].

\bibitem{Franciolini:2023agm}
G.~Franciolini, A.~Iovino, Junior., M.~Taoso and A.~Urbano,
Phys. Rev. D \textbf{109}, no.12, 123550 (2024)
doi:10.1103/PhysRevD.109.123550
[arXiv:2305.03491 [astro-ph.CO]].

\bibitem{Firouzjahi:2023ahg}
H.~Firouzjahi and A.~Riotto,
JCAP \textbf{02}, 021 (2024)
doi:10.1088/1475-7516/2024/02/021
[arXiv:2304.07801 [astro-ph.CO]].
	

\end{thebibliography}
\end{document}